\documentstyle[sprocl,epsf]{article}

\def\be{\begin{equation}}
\def\ee{\end{equation}}
\def\bea{\begin{eqnarray}}
\def\eea{\end{eqnarray}}

\begin{document}

\title{Instantons and the Large $N_c$ Limit of QCD}

\author{T.~Sch\"afer}

\address{Department of Physics, Duke University, Durham, NC 27708\\
Department of Physics, SUNY Stony Brook, Stony Brook, NY 11794\\ 
Riken-BNL Research Center, Brookhaven National Laboratory, 
Upton, NY 11973} 


\maketitle\abstracts{We summarize our current understanding
of instantons in the large $N_c$ limit of QCD. We also present
some recent results from simulations of the instanton liquid
in QCD for $N_c>3$.}

\section{Historical Overview}

  Both instantons and the large $N_c$ expansion are important
corner stones in our attempts to understand non-perturbative 
aspects of QCD. Instantons are related to topological aspects
of QCD and play and important role in understanding the $U(1)_A$
anomaly. There is also a growing body of evidence that the 
mechanism for chiral symmetry breaking in QCD is intimately 
connected with instantons. The large $N_c$ expansion is based
on the idea that $1/N_c$ can be used as expansion parameter 
in QCD. The main assumption is that QCD with many colors is 
a theory that exhibits confinement and chiral symmetry breaking, 
and that this theory is smoothly connected to the real world, 
in which $N_c$ is equal to three. There is an impressive amount 
of evidence, mainly of phenomenological nature, but also from 
the lattice, that this assumption is indeed correct. 

 It is commonly believed, however, that instantons and the 
large $N_c$ expansion are fundamentally incompatible with 
each other. This view is maybe best explained by providing
a short review of the literature on the subject. 
The instanton solution was discovered \cite{Belavin:fg} in 
1975. In the following year 't Hooft found the fermion zero 
mode in the background field of an instanton and explained
the connection between instantons and the $U(1)_A$ 
anomaly \cite{'tHooft:up}. In 1978 Witten wrote a very 
influential paper \cite{Witten:1978bc} entitled ``Instantons, 
the Quark Model, and the $1/N$ expansion''. He noted that 
straightforward $N_c$ counting suggests that the $\eta'$ mass 
squared scales as $1/N_c$, whereas classical effects, such as 
instantons, scale as $\exp(-1/g^2)\sim \exp(-N_c)$. He
explicitly stated that in trying to understand the $\eta'$
mass and other non-perturbative phenomena in QCD ``it is  
necessary to choose between the large $N$ expansion and 
instantons''. In 1979 Witten \cite{Witten:1979vv} and 
Veneziano \cite{Veneziano:1979ec} derived a relation 
between the mass of the $\eta'$ and the topological 
susceptibility in pure gauge theory,
\be 
\label{WV}
\frac{f_\pi^2}{2N_f} (m_\eta^2+m_{\eta'}^2-2m_K^2) = \chi_{top}.
\ee
Witten suggested that for all $N_c$ instantons ``melt''
and cannot be distinguished from perturbative fluctuations. He
also proposed that the topological susceptibility is of order
$O(1)$ in the large $N_c$ limit, and that it is dominated by
non-perturbative fluctuations other than instantons, 
presumably related to confinement. Indeed, Veneziano's paper 
is entitled ``$U(1)$ without Instantons''. 

  Simple $N_c$ counting implies that both the $\eta'-\pi$ 
splitting and the $\rho-\omega$ splitting are of order $1/N_c$.
In the real world, of course, these two phenomena are very 
different in magnitude. This suggests that there is some 
physical effect in the $\eta'$ channel that leads to unusually
large $1/N_c$ corrections. This observation was discussed in
more detail by Novikov \cite{Novikov:xj} et al.~in a paper 
entitled ``Are all hadrons alike?''. Novikov et al.~observed 
that large corrections to the OZI rule appear many channels, 
not just pseudoscalar mesons, but also scalar mesons and 
glueballs. They also noticed that these are precisely the 
same channels in which direct instanton effects appear. 

  This observation led to the formulation of the 
instanton liquid model of the QCD vacuum 
\cite{Shuryak:1981ff,Diakonov:1983hh,Schafer:1996wv}.
The instanton liquid model postulates that the $N_c=3$
QCD vacuum is populated by localized, approximately 
dual or anti-self dual lumps, instantons. The density
of instantons is approximately $(N/V)\simeq 1\, {\rm 
fm}^{-4}$ while the size is $\rho\simeq 1/3\, {\rm fm}$.
These numbers reproduce the topological susceptibility
in the pure gauge theory $\chi_{top}\simeq (200\, {\rm 
MeV})^4$ and the chiral condensate $\langle \bar{\psi}
\psi\rangle \simeq -(230\,{\rm MeV})^3$. More detailed
calculations show that the instanton liquid model 
successfully describes an impressive amount of date 
on hadronic correlation functions 
\cite{Shuryak:1993,Schafer:1996wv,Schafer:2000rv}.
 
  In the 1980's researchers also began to study the 
topological structure of QCD on the lattice. It was
found that the topological susceptibility in pure 
gauge QCD is \cite{Teper:1999wp} $\chi_{top}\simeq (200\, 
{\rm MeV})^4$, as predicted by the Witten-Veneziano 
relation equ.~(\ref{WV}). However, it was also observed
that the topological susceptibility is very stable
under cooling, and appears to be dominated by semi-classical
configurations. Lattice simulations also appear to 
confirm the values of the key parameters of the instanton
liquid \cite{Chu:1994},  $(N/V)\simeq 1\, {\rm fm}^{-4}$ and 
$\rho\simeq 1/3\, {\rm fm}$. 

  More recently lattice simulations have started to 
focus on the structure of low-lying eigenvectors of 
the Dirac operator and the mechanism of chiral symmetry
breaking. The instanton model predicts that the lowest 
eigenstates of the Dirac operator, which dominate chiral
symmetry breaking, are linear combinations of localized, 
approximately chiral states associated with the fermionic
zero modes of individual instantons and anti-instantons
\cite{Diakonov:1985}. This picture has been confirmed by 
lattice calculations \cite{DeGrand:2000gq,Horvath:2001ir}, 
although there is some controversy concerning the question 
whether the size of the chiral lumps is in agreement with 
the instanton prediction \cite{DeGrand:2000gq,Dong:2001kn}.

\section{Instantons and the OZI rule}

  In this section we would like to remind the reader how instantons
lead to large violations of the OZI rule. 't Hooft\cite{'tHooft:up} 
explained that the effect of an instanton on fermionic correlation 
functions can be summarized in terms of an effective interaction
\begin{equation} 
L = G \left[
  (\bar{\psi}\tau_a\psi)^2 - (\bar{\psi}\psi)^2 
- (\bar{\psi}i\gamma_5\tau_a\psi)^2 + (\bar{\psi}i\gamma_5\psi)^2 \right],
\end{equation}
where $G$ is related to the tunneling amplitude and $\tau$ is 
an isospin matrix. On the single instanton level we can directly
read off the interaction in the channel characterized by the current 
$\bar\psi\Gamma\psi$. The interaction is attractive in the pion 
$(\bar\psi i\gamma_5\vec{\tau}\psi)$ and sigma $(\bar\psi\psi)$
channels, and repulsive in the eta prime $(\bar\psi i\gamma_5\psi)$
and $a_0$ $(\bar\psi\vec{\tau}\psi)$ channels. We note that the 
OZI violating interaction the $\eta'$ and $\sigma$ channel is
not suppressed with respect to the OZI allowed contribution 
in the $\pi$ and $a_0$ sector. 

 It is worth repeating why this is so. The instanton 
interaction corresponds to the contribution of fermionic zero
modes to the quark propagator. Since there is exactly one zero
mode for every flavor, there are no diagonal $(\bar{u}u)(\bar{u}u)$ 
or $(\bar{d}d)(\bar{d}d)$ interactions. Second, since the fermion
zero modes for quarks and anti-quarks have opposite chirality,
the interaction is also off-diagonal in the basis spanned by 
right and left-handed fermions $q_R,\,q_L$. As a consequence
the sign of the interaction flips in going from the scalar to 
the pseudoscalar channel, and in going from $I=0$ to $I=1$ states.
Also, to leading order there is no instanton contribution 
to the interaction in vector channels. 

 In order to be more quantitative we consider correlation 
functions involving the scalar and pseudoscalar currents 
introduced above. The correlators are defined by
\bea
\Pi_\pi(x,y) &=& \langle {\rm Tr}\left(
 S(x,y)i\gamma_5 S(y,x)i\gamma_5\right)\rangle, \\
\Pi_\delta(x,y) &=& \langle {\rm Tr}\left(
 S(x,y) S(y,x)\right)\rangle ,\\
\Pi_{\eta'\!}(x,y) &=& \langle {\rm Tr}\left(
 S(x,y)i\gamma_5 S(y,x)i\gamma_5\right)\rangle 
 - 2 \langle {\rm Tr}\left(i\gamma_5 S(x,x)\right)
           {\rm Tr}\left(i\gamma_5 S(y,y)\right)\rangle ,\\
\Pi_{\sigma}(x,y) &=& \langle {\rm Tr}\left(
 S(x,y) S(y,x)\right)\rangle 
 - 2 \langle {\rm Tr}\left(S(x,x)\right)
           {\rm Tr}\left(S(y,y)\right)\rangle ,
\eea
where $S(x,y)$ is the fermion propagator and $\langle.\rangle$ 
denotes the average over all gauge configurations. The OZI-violating
difference between the $\pi,\eta'$ and $\sigma,a_0$ channels is
determined by ``disconnected'' (or double hairpin) contributions 
to the correlation functions. 

\begin{figure}
\begin{center}
\vspace*{-1.2cm}
\epsfxsize=8cm
\epsffile{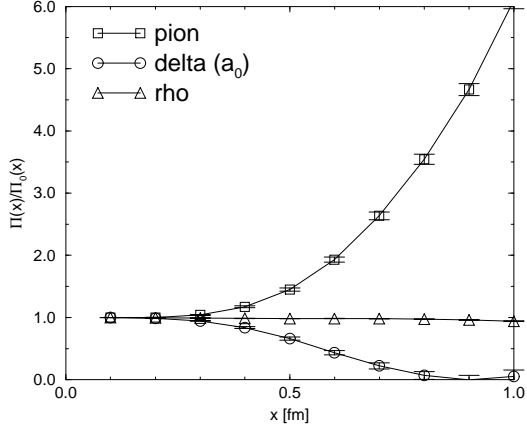}
\vspace*{0cm}
\end{center}
\caption{Correlation functions in the pion, delta ($a_0$), 
and rho meson channels. The correlators are normalized 
to free field behavior ($\Pi_0(x)\sim x^{-6}$).}
\end{figure}
\begin{figure}
\begin{center}
\vspace*{-1.2cm}
\epsfxsize=8cm
\epsffile{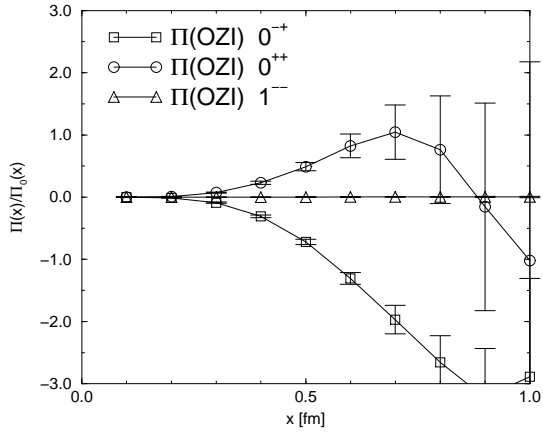}
\vspace*{0cm}
\end{center}
\caption{OZI violating disconnected correlation functions 
in the pseudoscalar ($\eta'-\pi$), scalar ($\sigma-a_0$),
and vector ($\omega-\rho$) channel. The correlators are 
normalized as in Fig.1.}
\end{figure}

 In Figs. 1 and 2 we show results for the correlation functions 
obtained in unquenched instanton simulations of the instanton
liquid \cite{Schafer:1996uz}. We observe that the correlation
function in the OZI violating $\eta'-\pi$ and $\sigma-a_0$
channels is as large as the correlation function in the 
OZI allowed $\pi$ channel whereas the OZI violating $\omega-
\rho$ channel is much smaller. This pattern was also 
found in a lattice calculation by Isgur and Thacker 
\cite{Isgur:2000ts,Schafer:2000hn}. It would be interesting 
to perform similar calculations in other channels in
which large violations of the OZI rule might be present.
Examples are the mixing of scalar glueballs with scalar
mesons or multi-pion states, the $\Delta I=1/2$ rule, and
OZI suppressed decays of heavy quark bound states. 

\section{Instantons at Large $N_c$}

 In the previous section we argued that the OZI violating 
direct instanton contribution in the $\eta'$ channel is not
suppressed compared to the instanton contribution in the 
$\pi$ channel. But what is the $N_c$ dependence of these
effects? Witten argued that instanton effects are proportional
to $\exp(-8\pi^2/g^2)$ and therefore scale as $\exp(-N_c)$. 
But this is only true for instantons at the cutoff scale. 
In perturbation theory, the tunneling rate is of the form
\be
 d n(\rho) \sim \rho^{-5}(\rho\Lambda)^b \,d\rho ,
\ee
where $b=11N_c/3-2N_f/3$ is the first coefficient of the 
beta function. This implies that small instantons $\rho
<\Lambda^{-1}$ are suppressed as $N_c\to \infty$, but 
large instantons are not. 

 This is observation is not very useful, because instantons 
of size $\Lambda^{-1}$ have action $S\sim 1$ and are clearly 
not semi-classical objects. We can be somewhat more 
quantitative, however. The one-loop result contains
additional $N_c$ dependent factors that are related 
to the collective coordinate measure. Exponentiating everything 
we can write \cite{Teper:1979tq,Neuberger:1980as,Shuryak:1995pv}
\be 
\label{n(rho)}
\frac{dN}{d\rho} \sim \exp[ N_c F(\rho) ],
\ee
with $F(\rho)=2-s(\rho)+2\log(s(\rho))+\ldots$ and $s(\rho)
\equiv S(\rho)/N_c = (8\pi^2)/(N_c g^2(\rho))$. The function 
$F(\rho)$ has a non-trivial zero for $s^*\sim 5$ so the density 
of instantons of size $\rho=\rho^* \sim 0.2\Lambda^{-1}$ fm is 
fixed as $N_c\to\infty$. This is a much more optimistic conclusions, 
because it implies that instantons with action $S=N_c s^* 
\sim 5N_c \gg 1$ can survive in the large $N_c$ limit. The 
precise value of $\rho^*$ and $s^*$ depends on the renormalization 
scheme and higher order corrections. A fixed point in the instanton 
size distribution was indeed observed in the lattice calculation 
of Lucini and Teper \cite{Lucini:2001ej}, see also Fig.~3. However, 
these authors also find that the density of instantons of size 
$\rho>\rho^*$ keeps growing as $N_c\to\infty$, and that as a result 
there is no well defined average instanton size.

\section{Supersymmetric Theories}

  While the fate of instantons in the large $N_c$ limit of QCD
is still mysterious some progress has been made in understanding
instantons in the large $N_c$ limit of supersymmetric gauge 
theories. The most impressive accomplishment is Maldacena's
discovery of an explicit master field in the case of $N=4$ 
supersymmetric QCD at large $N_c$ in the limit of large 
't Hooft coupling $\lambda=g^2 N_c$. The master field is 
given by supergravity in an $AdS_5\times S_5$ background
together with a set of rules that relate gauge theory to 
supergravity observables. It is very interesting to study 
how instantons appear in this correspondence \cite{Bianchi:1998nk}.
On the supergravity side gauge theory instantons appear as
D-instantons and are characterized by a point on $AdS_5
\times S_5$. This integration over this point corresponds
to the collective coordinate integration on the gauge 
theory side. The $AdS_5$ part is related to the integration
over $d^4xd\rho/\rho^5$ while the $S_5$ arises from the
fermion zero modes \cite{Dorey:1998xe}.

 The case of multi-instanton configurations is even more 
interesting. Naively one would think that a $k$-instanton
contribution involves an integration over $(AdS_5\times
S_5)^k$. In the large $N_c$ limit one finds, however, 
that the saddle point configuration is given by $k$
instantons in commuting $SU(2)$ subgroups of $SU(N_c)$
all with the same size and located at the same point 
\cite{Dorey:1999pd}. As a result, the collective
coordinate measure contains only one copy of $AdS_5
\times S_5$. The multi-instanton configuration is 
bound by fermion exchanges. 

 Progress was also made in understanding $N=2$ SUSY QCD.
Seiberg and Witten determined the low energy effective action of 
this theory \cite{Seiberg:1994rs}. In the semi-classical limit, 
the result can be expressed as the perturbative one-loop 
contribution plus an infinite series on $k$-instanton corrections, 
and, remarkably, nothing else. This would seem to imply that 
the large $N_c$ limit of this theory is rather boring. Instantons
are suppressed as $\exp(-N_c)$, monopoles have masses $O(N_c)$, 
and one is left with only the perturbative part. 

 This is not correct, however. The methods of Seiberg and Witten
were generalized to arbitrary $N_c$ by Klemm et al.~\cite{Klemm:1994qs}. 
This result was analyzed in more detail by Douglas and Shenker 
\cite{Douglas:1995nw}. The moduli space of the theory is 
characterized by $N_c-1$ Higgs expectation values. At most
points on the moduli space the large $N_c$ limit is indeed
trivial but Douglas and Shenker identified a special form 
of the large $N_c$ limit in which instantons and monopoles
survive.

\section{The Instanton Liquid at Large $N_c$}

 In this section we wish to study the question whether it is 
possible to construct a large $N_c$ instanton ensemble in QCD
that is consistent with standard large $N_c$ counting. 
Equ.~(\ref{n(rho)}) shows that instantons of action $S\sim 
N_c\gg 1$ and size $\rho \sim N_c^0$ can survive in the 
large $N_c$ limit. It is quite natural to assume that the 
total density of these objects would scale as $(N/V) \sim 
N_c$ because instantons are essentially $SU(2)$ configurations
and the number of mutually commuting $SU(2)$ subgroups of 
$SU(N_c)$ grows as $N_c$. This means that we can have $O(N_c)$ 
instantons and the instanton liquid remains dilute. Using
the trace anomaly relation
\be 
 \langle T_{\mu\mu}\rangle = -\frac{b}{32\pi^2}
  \langle g^2 G^a_{\mu\nu} G^a_{\mu\nu} \rangle 
\ee
with $b=11N_c/2-2N_f/2$ we also see that the instanton 
contribution to the vacuum energy scales as $\epsilon\sim N_c^2$. 
This is identical to the scaling behavior of the perturbative 
part.

\begin{figure}
\begin{center}
\vspace*{0cm}
\epsfxsize=8cm
\epsffile{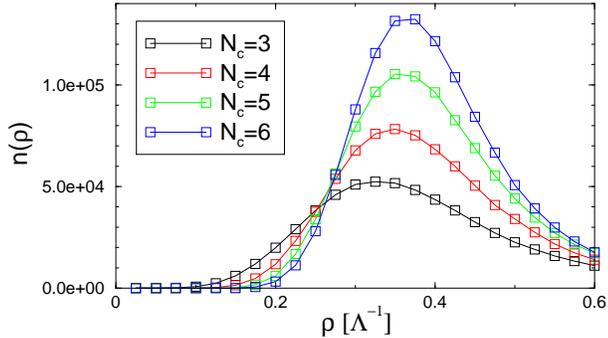}
\vspace*{0cm}
\end{center}
\caption{Instanton size distribution for different values of $N_c$
obtained from a numerical simulation of the interacting instanton
liquid. The instanton size is given in units of the Pauli-Vilars
scale parameter $\Lambda\simeq 200$ MeV.}
\end{figure}

\begin{figure}
\begin{center}
\vspace*{0cm}
\epsfxsize=8cm
\epsffile{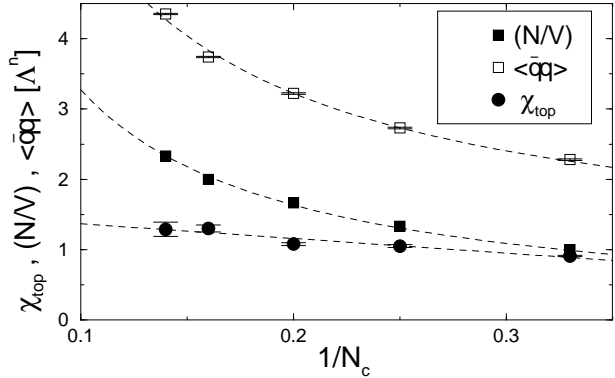}
\vspace*{0cm}
\end{center}
\caption{Instanton density $(N/V)$, chiral condensate $\langle
\bar{q}q\rangle$ and topological susceptibility obtained from 
a numerical simulation of the interacting instanton liquid.
All quantities are given in units of appropriate powers of
the Pauli-Vilars scale parameter.}
\end{figure}

  In order to construct a consistent ensemble of instantons 
we have to make an assumption concerning the fate of large 
instantons. In the following we shall assume that there is
a classical $O(S_0) = O(N_c)$ core in the instanton 
interaction for instantons that overlap in group space. 
This core excludes configurations with large or strongly
overlapping instantons that are not semi-classical. The 
parameters of the core were fitted to reproduce the 
phenomenological values of the instanton size and density
for \cite{Schafer:1996wv} $N_c=3$. Instanton size distributions
obtained from numerical simulations of the instanton liquid
for $N_c=3,\ldots, 6$ are shown in Fig.~3. In these simulations
we have assumed that $(N/V)\sim N_c$, but this assumption
can be verified by computing the free energy of the instanton 
liquid as a function of \cite{Schafer:2010} $(N/V)$. We note 
that the distributions show the fixed point discussed above 
and observed in the lattice simulations of Lucini and Teper. 

 We have also computed the quark condensate and the 
topological susceptibility in pure gauge theory, see Fig.~4. 
We observe that the quark condensate scales as $\langle\bar{q}
q\rangle \sim N_c$. This scaling can be understood using random 
matrix arguments. The quark condensate is approximately 
given by
\be 
\langle \bar{q}q\rangle = -\frac{1}{\pi\rho}
 \left( \frac{3N_c}{2}\frac{N}{V}\right)^{1/2}.
\ee
The total number of states in the zero-mode zone
scales as $N_{zmz}\sim (N/V)\sim N_c$. This contributes a
factor $N_c^{1/2}$ to the scaling behavior of $\langle
\bar{q}q\rangle$. The second factor of $N_c^{1/2}$ 
arises from the fact that the quark condensate is 
inversely proportional to the average matrix element
$|T_{IA}|$ of the Dirac operator between fermion zero 
modes. Since instanton zero modes live in $SU(2)$
subgroups $|T_{IA}|^2$ scales as $1/N_c$. Similar 
arguments can be used to show that the pion decay 
constant scales as $f_\pi^2\sim N_c$. 

 For a dilute gas of instantons we would expect that 
the topological susceptibility in pure gauge theory
scales as $\chi_{top}=\langle Q^2_{top}\rangle /V
\simeq (N/V) \sim N_c$, contrary to the standard large 
$N_c$ assumption $\chi_{top}\sim N_c^0$. Further 
support for this assumption was recently provided
by Witten \cite{Witten:1998uk}. We have to keep in 
mind, however, that the result $\chi_{top}\simeq
(N/V)$ is based on the idea that topological charge 
fluctuations are Poissonian. Since the classical 
interaction between instantons also grows with $N_c$,
this is not necessarily the case \cite{Carter:2001ih}. 
Indeed, simulations carried out for $N_c=3,\ldots,7$ 
seem to indicate that the topological susceptibility 
remains finite in the large $N_c$ limit, see Fig.~4.

\section{Instead of Conclusions}

  We have argued that instantons provide a very successful
explanation of the pattern of OZI violation observed in QCD
for $N_c=3$ colors. The fate of instantons in the large $N_c$ 
limit remains unclear. It may well be that instantons become 
large and overlap strongly, and the semi-classical description 
breaks down completely. However, it is also possible that 
the semi-classical picture survives. This would seem to require
large cancellations and fine tuning, but for reasons that 
we do not understand the parameters of the instanton liquid 
in QCD with $N_c=3$ colors are already remarkably close 
to the critical point where a smooth large $N_c$ limit 
is possible.


\section*{References}
\begin{thebibliography}{99}

\bibitem{Belavin:fg}
A.~A.~Belavin, A.~M.~Polyakov, A.~S.~Shvarts and Y.~S.~Tyupkin,
Phys.\ Lett.\ B {\bf 59}, 85 (1975).

\bibitem{'tHooft:up}
G.~'t Hooft,
Phys.\ Rev.\ Lett.\ {\bf 37}, 8 (1976).

\bibitem{Witten:1978bc}
E.~Witten,
Nucl.\ Phys.\ B {\bf 149}, 285 (1979).

\bibitem{Witten:1979vv}
E.~Witten,
Nucl.\ Phys.\ B {\bf 156}, 269 (1979).

\bibitem{Veneziano:1979ec}
G.~Veneziano,
Nucl.\ Phys.\ B {\bf 159}, 213 (1979).

\bibitem{Novikov:xj}
V.~A.~Novikov, M.~A.~Shifman, A.~I.~Vainshtein and V.~I.~Zakharov,
Nucl.\ Phys.\ B {\bf 191}, 301 (1981).

\bibitem{Shuryak:1981ff}
E.~V.~Shuryak,
Nucl.\ Phys.\ B {\bf 203}, 93 (1982).

\bibitem{Diakonov:1983hh}
D.~Diakonov and V.~Y.~Petrov,
Nucl.\ Phys.\ B {\bf 245}, 259 (1984).

\bibitem{Schafer:1996wv} 
T.~Sch{\"a}fer and E.~V.~Shuryak, 
Rev.\ Mod.\ Phys.\ {\bf 70}, 323 (1998)
[hep-ph/9610451].

\bibitem{Shuryak:1993}
E.~V.~Shuryak,
Rev.\ Mod.\ Phys.\  {\bf 65}, 1 (1993).

\bibitem{Schafer:2000rv}
T.~Sch\"afer and E.~V.~Shuryak,
Phys.\ Rev.\ Lett.\ {\bf 86}, 3973 (2001)
[hep-ph/0010116].

\bibitem{Teper:1999wp}
M.~Teper,
Nucl.\ Phys.\ Proc.\ Suppl.\  {\bf 83}, 146 (2000)
[hep-lat/9909124].

\bibitem{Chu:1994}
M.~C.~Chu, G.~M.~Grandy, S.~Huang, and J.~W.~Negele,
Phys.\ Rev.\ D {\bf 49} 6039 (1994).
[hep-lat/9312071].

\bibitem{Diakonov:1985}
D.~Diakonov and V.~Y.~Petrov,
Sov.\ Phys.\ JETP {\bf 62} 204 (1985).

\bibitem{DeGrand:2000gq}
T.~DeGrand and A.~Hasenfratz,
Phys.\ Rev.\ D {\bf 64}, 034512 (2001)
[hep-lat/0012021].

\bibitem{Horvath:2001ir}
I.~Horvath, N.~Isgur, J.~McCune and H.~B.~Thacker,
Phys.\ Rev.\ D {\bf 65}, 014502 (2002)
[hep-lat/0102003].
T.~DeGrand and A.~Hasenfratz,
Phys.\ Rev.\ D {\bf 65}, 014503 (2002)
[hep-lat/0103002].
I.~Hip, T.~Lippert, H.~Neff, K.~Schilling and W.~Schroers,
Phys.\ Rev.\ D {\bf 65}, 014506 (2002)
[hep-lat/0105001].
R.~G.~Edwards and U.~M.~Heller,
Phys.\ Rev.\ D {\bf 65}, 014505 (2002)
[hep-lat/0105004].
T.~Blum {\it et al.},
Phys.\ Rev.\ D {\bf 65}, 014504 (2002)
[hep-lat/0105006].
C.~Gattringer, M.~Gockeler, P.~E.~Rakow, S.~Schaefer and A.~Schafer,
Nucl.\ Phys.\ B {\bf 617}, 101 (2001)
[hep-lat/0107016].

\bibitem{Dong:2001kn}
S.~J.~Dong {\it et al.},
Nucl.\ Phys.\ Proc.\ Suppl.\  {\bf 106}, 563 (2002)
[hep-lat/0110037].

\bibitem{Schafer:1996uz}
T.~Sch{\"a}fer and E.~V.~Shuryak,
Phys.\ Rev.\  {\bf D54}, 1099 (1996)
[hep-ph/9512384].

\bibitem{Isgur:2000ts}
N.~Isgur and H.~B.~Thacker,
Phys.\ Rev.\ D {\bf 64}, 094507 (2001)
[hep-lat/0005006].

\bibitem{Schafer:2000hn}
T.~Schafer and E.~V.~Shuryak,
hep-lat/0005025.

\bibitem{Teper:1979tq}
M.~J.~Teper,
Z.\ Phys.\ C {\bf 5}, 233 (1980).

\bibitem{Neuberger:1980as}
H.~Neuberger,
Phys.\ Lett.\ B {\bf 94}, 199 (1980).

\bibitem{Shuryak:1995pv}
E.~V.~Shuryak,
Phys.\ Rev.\ D {\bf 52}, 5370 (1995)
[hep-ph/9503467].

\bibitem{Lucini:2001ej}
B.~Lucini and M.~Teper,
JHEP {\bf 0106}, 050 (2001)
[hep-lat/0103027].

\bibitem{Maldacena:1997re}
J.~Maldacena,
Adv.\ Theor.\ Math.\ Phys.\  {\bf 2}, 231 (1998)
[hep-th/9711200].

\bibitem{Bianchi:1998nk}
M.~Bianchi, M.~B.~Green, S.~Kovacs and G.~Rossi,
JHEP {\bf 9808}, 013 (1998)
[hep-th/9807033].

\bibitem{Dorey:1998xe}
N.~Dorey, V.~V.~Khoze, M.~P.~Mattis and S.~Vandoren,
Phys.\ Lett.\ B {\bf 442}, 145 (1998)
[hep-th/9808157].

\bibitem{Dorey:1999pd}
N.~Dorey, T.~J.~Hollowood, V.~V.~Khoze, M.~P.~Mattis and S.~Vandoren,
Nucl.\ Phys.\ B {\bf 552}, 88 (1999)
[hep-th/9901128].

\bibitem{Seiberg:1994rs}
N.~Seiberg and E.~Witten,
Nucl.\ Phys.\ B {\bf 426}, 19 (1994)
[hep-th/9407087].

\bibitem{Klemm:1994qs}
A.~Klemm, W.~Lerche, S.~Yankielowicz and S.~Theisen,
Phys.\ Lett.\ B {\bf 344}, 169 (1995)
[hep-th/9411048].

\bibitem{Douglas:1995nw}
M.~R.~Douglas and S.~H.~Shenker,
Nucl.\ Phys.\ B {\bf 447}, 271 (1995)
[hep-th/9503163].

\bibitem{Schafer:2010}
T.~Sch\"afer, in preparation.

\bibitem{Witten:1998uk}
E.~Witten,
Phys.\ Rev.\ Lett.\  {\bf 81}, 2862 (1998)
[hep-th/9807109].

\bibitem{Carter:2001ih}
G.~W.~Carter and E.~V.~Shuryak,
Phys.\ Lett.\ B {\bf 524}, 297 (2002).


\end{thebibliography}
\end{document}